\documentclass[preprint2]{aastex}

\usepackage{amsmath}
\usepackage{amsfonts}
\usepackage{amssymb}

\begin{document}
\title{SEARCH FOR GAMMA-RAYS FROM THE UNUSUALLY BRIGHT GRB~130427A WITH THE HAWC GAMMA-RAY OBSERVATORY}

\author{A.~U.~Abeysekara\altaffilmark{1,6},
R.~Alfaro\altaffilmark{2,7},
C.~Alvarez\altaffilmark{3},
J.~D.~{\'A}lvarez\altaffilmark{4},
R.~Arceo\altaffilmark{3},
J.~C.~Arteaga-Vel{\'a}zquez\altaffilmark{4},
H.~A.~Ayala~Solares\altaffilmark{5},
A.~S.~Barber\altaffilmark{6},
B.~M.~Baughman\altaffilmark{7},
N.~Bautista-Elivar\altaffilmark{8},
S.~Y.~BenZvi\altaffilmark{9,15},
M.~Bonilla~Rosales\altaffilmark{10},
J.~Braun\altaffilmark{7,15},
K.~S.~Caballero-Mora\altaffilmark{11},
A.~Carrami{\~n}ana\altaffilmark{10},
M.~Castillo\altaffilmark{12},
U.~Cotti\altaffilmark{4},
J.~Cotzomi\altaffilmark{12},
E.~de~la~Fuente\altaffilmark{13},
C.~De~Le{\'o}n\altaffilmark{4},
T.~DeYoung\altaffilmark{1},
R.~Diaz~Hernandez\altaffilmark{10},
B.~L.~Dingus\altaffilmark{14},
M.~A.~DuVernois\altaffilmark{15},
R.~W.~Ellsworth\altaffilmark{16,7},
D.~W.~Fiorino\altaffilmark{15},
N.~Fraija\altaffilmark{17},
A.~Galindo\altaffilmark{10},
F.~Garfias\altaffilmark{17},
M.~M.~Gonz{\'a}lez\altaffilmark{17,7},
J.~A.~Goodman\altaffilmark{7},
M.~Gussert\altaffilmark{18},
Z.~Hampel-Arias\altaffilmark{15},
J.~P.~Harding\altaffilmark{14},
P.~H{\"u}ntemeyer\altaffilmark{5},
C.~M.~Hui\altaffilmark{5},
A.~Imran\altaffilmark{14,15},
A.~Iriarte\altaffilmark{17},
P.~Karn\altaffilmark{15,27},
D.~Kieda\altaffilmark{6},
G.~J.~Kunde\altaffilmark{14},
A.~Lara\altaffilmark{19},
R.~J.~Lauer\altaffilmark{20},
W.~H.~Lee\altaffilmark{17},
D.~Lennarz\altaffilmark{21,*},
H.~Le{\'o}n~Vargas\altaffilmark{2},
J.~T.~Linnemann\altaffilmark{1},
M.~Longo\altaffilmark{18},
R.~Luna-Garc{\'\i}a\altaffilmark{23},
K.~Malone\altaffilmark{22},
A.~Marinelli\altaffilmark{2},
S.~S.~Marinelli\altaffilmark{1},
H.~Martinez\altaffilmark{11},
O.~Martinez\altaffilmark{12},
J.~Mart{\'\i}nez-Castro\altaffilmark{23},
J.~A.~J.~Matthews\altaffilmark{20},
E.~Mendoza~Torres\altaffilmark{10},
P.~Miranda-Romagnoli\altaffilmark{24},
E.~Moreno\altaffilmark{12},
M.~Mostaf{\'a}\altaffilmark{22},
L.~Nellen\altaffilmark{25},
M.~Newbold\altaffilmark{6},
R.~Noriega-Papaqui\altaffilmark{24},
T.~O.~Oceguera-Becerra\altaffilmark{13,2},
B.~Patricelli\altaffilmark{17},
R.~Pelayo\altaffilmark{23},
E.~G.~P{\'e}rez-P{\'e}rez\altaffilmark{8},
J.~Pretz\altaffilmark{22},
C.~Rivi\`ere\altaffilmark{17,7},
D.~Rosa-Gonz{\'a}lez\altaffilmark{10},
H.~Salazar\altaffilmark{12},
F.~Salesa~Greus\altaffilmark{22},
A.~Sandoval\altaffilmark{2},
M.~Schneider\altaffilmark{26},
G.~Sinnis\altaffilmark{14},
A.~J.~Smith\altaffilmark{7},
K.~Sparks~Woodle\altaffilmark{22},
R.~W.~Springer\altaffilmark{6},
I.~Taboada\altaffilmark{21},
K.~Tollefson\altaffilmark{1},
I.~Torres\altaffilmark{10},
T.~N.~Ukwatta\altaffilmark{1,14},
L.~Villase{\~n}or\altaffilmark{4},
T.~Weisgarber\altaffilmark{15},
S.~Westerhoff\altaffilmark{15},
I.~G.~Wisher\altaffilmark{15},
J.~Wood\altaffilmark{7},
G.~B.~Yodh\altaffilmark{27},
P.~W.~Younk\altaffilmark{14},
D.~Zaborov\altaffilmark{22},
A.~Zepeda\altaffilmark{11},
H.~Zhou\altaffilmark{5},
\newline {(The HAWC collaboration)}}

\altaffiltext{1}{Department of Physics \& Astronomy, Michigan State University, East Lansing, MI, USA}
\altaffiltext{2}{Instituto de F{\'\i}sica, Universidad Nacional Aut{\'o}noma de M{\'e}xico, M{\'e}xico D.F., Mexico}
\altaffiltext{3}{CEFyMAP, Universidad Aut{\'o}noma de Chiapas, Tuxtla Guti{\'e}rrez, Chiapas, Mexico}
\altaffiltext{4}{Universidad Michoacana de San Nicol{\'a}s de Hidalgo, Morelia, Michoac{\'a}n, Mexico}
\altaffiltext{5}{Department of Physics, Michigan Technological University, Houghton, MI, USA}
\altaffiltext{6}{Department of Physics \& Astronomy, University of Utah, Salt Lake City, UT, USA}
\altaffiltext{7}{Department of Physics, University of Maryland, College Park, MD, USA}
\altaffiltext{8}{Universidad Polit{\'e}cnica de Pachuca, Municipio de Zempoala, Hidalgo, Mexico}
\altaffiltext{9}{Department of Physics \& Astronomy, University of Rochester, Rochester, NY, USA}
\altaffiltext{10}{Instituto Nacional de Astrof{\'\i}sica, {\'O}ptica y Electr{\'o}nica, Tonantzintla, Puebla, Mexico}
\altaffiltext{11}{Centro de Investigaci{\'o}n y de Estudios Avanzados del Instituto Polit{\'e}cnico Nacional, M{\'e}xico D.F., Mexico}
\altaffiltext{12}{Facultad de Ciencias F{\'\i}sico Matem{\'a}ticas, Benem{\'e}rita Universidad Aut{\'o}noma de Puebla, Ciudad Universitaria, Puebla, Mexico}
\altaffiltext{13}{Departamento de F{\'\i}sica, Centro Universitario de Ciencias Exactas e Ingenier{\'\i}as, Universidad de Guadalajara, Guadalajara, Mexico}
\altaffiltext{14}{Physics Division, Los Alamos National Laboratory, Los Alamos, NM, USA}
\altaffiltext{15}{Wisconsin IceCube Particle Astrophysics Center (WIPAC) and Department of Physics, University of Wisconsin-Madison, Madison, WI, USA}
\altaffiltext{16}{School of Physics, Astronomy \& Computational Sciences, George Mason University, Fairfax, VA, USA}
\altaffiltext{17}{Instituto de Astronom{\'\i}a, Universidad Nacional Aut{\'o}noma de M{\'e}xico, M{\'e}xico D.F., Mexico}
\altaffiltext{18}{Physics Department, Colorado State University, Fort Collins, CO, USA}
\altaffiltext{19}{Instituto de Geof{\'\i}sica, Universidad Nacional Aut{\'o}noma de M{\'e}xico, M{\'e}xico D.F., Mexico}
\altaffiltext{20}{Department of Physics \& Astronomy, University of New Mexico, Albuquerque, NM, USA}
\altaffiltext{21}{School of Physics and Center for Relativistic Astrophysics, Georgia Institute of Technology, Atlanta, GA, USA}
\altaffiltext{22}{Department of Physics, Pennsylvania State University, University Park, PA, USA}
\altaffiltext{23}{Centro de Investigaci{\'o}n en Computaci{\'o}n, Instituto Polit{\'e}cnico Nacional, M{\'e}xico D.F., Mexico}
\altaffiltext{24}{Universidad Aut{\'o}noma del Estado de Hidalgo, Pachuca, Hidalgo, Mexico}
\altaffiltext{25}{Instituto de Ciencias Nucleares, Universidad Nacional Aut{\'o}noma de M{\'e}xico, M{\'e}xico D.F., Mexico}
\altaffiltext{26}{Santa Cruz Institute for Particle Physics, University of California, Santa Cruz, Santa Cruz, CA, USA}
\altaffiltext{27}{Department of Physics \& Astronomy, University of California, Irvine, Irvine, CA, USA}
\altaffiltext{*}{dirk.lennarz@gatech.edu}

\shorttitle{Search for VHE Emission from GRB~130427A with HAWC}
\shortauthors{HAWC Collaboration et al.}
\authoraddr{D.~Lennarz, \email{dirk.lennarz@gatech.edu}}

\begin{abstract}
The first limits on the prompt emission from the long gamma-ray burst (GRB) 130427A in the $>100\nobreakspace\rm{GeV}$ energy band are reported. GRB~130427A was the most powerful burst ever detected with a redshift $z\lesssim0.5$ and featured the longest lasting emission above $100\nobreakspace\rm{MeV}$. The energy spectrum extends at least up to $95\nobreakspace\rm{GeV}$, clearly in the range observable by the High Altitude Water Cherenkov (HAWC) Gamma-ray Observatory, a new extensive air shower detector currently under construction in central Mexico. The burst occurred under unfavourable observation conditions, low in the sky and when HAWC was running 10\% of the final detector. Based on the observed light curve at MeV--GeV energies, eight different time periods have been searched for prompt and delayed emission from this GRB. In all cases, no statistically significant excess of counts has been found and upper limits have been placed. It is shown that a similar GRB close to zenith would be easily detected by the full HAWC detector, which will be completed soon. The detection rate of the full HAWC detector may be as high as one to two GRBs per year. A detection could provide important information regarding the high energy processes at work and the observation of a possible cut-off beyond the \emph{Fermi}-LAT energy range could be the signature of gamma-ray absorption, either in the GRB or along the line of sight due to the extragalactic background light.
\end{abstract}

\keywords{gamma-ray burst: individual (GRB\nobreakspace130427A) --- gamma rays: general}

\section{Introduction}
Gamma-ray bursts (GRBs) are the most luminous objects known \citep[for a review see, e.g.][]{GRB_review} and although they have been studied since the late 1960s, their particle acceleration mechanisms are still poorly understood. The general picture is that a central engine, e.g. the core-collapse of a rapidly rotating star \citep[collapsar;][]{bib:collapsar} or the merger of two compact stellar remnants \citep{bib:short_GRB_origin_1,bib:short_GRB_origin_2}, creates a collimated relativistic outflow \citep[fireball; for a review see, e.g.][]{bib:Piran_fireball_review}. Internal shocks arise if relativistic jets with varied Lorentz factors collide in the outflow, and once the outflowing material interacts with the surrounding material it creates external shocks. The main GRB lasts from $10^{-2}\nobreakspace\rm{s}$ to $10^{3}\nobreakspace\rm{s}$ and most of the energy is emitted in the keV--MeV range. This prompt emission might arise from internal shocks or photospheric emission. It is followed by a multi-wavelength afterglow that lasts significantly longer and is generally attributed to the external shocks.

The Large Area Telescope (LAT) on board the \emph{Fermi Gamma-Ray Space Telescope} (\emph{Fermi}-LAT) recently found that the $>100\nobreakspace\rm{MeV}$ emission of GRBs not only starts later than the keV--MeV emission (e.g. reaching delays up to $40\nobreakspace\rm{s}$ for GRB\nobreakspace090626), but is also temporally extended \citep{bib:Fermi_LAT_GRB_catalogue}. In the afterglow synchrotron model, electrons accelerated by the external shock produce the temporally extended GeV emission via synchrotron radiation. However, in this scenario the maximum photon energy is limited \citep[e.g.][]{bib:max_synchrotron}. Hence, the observation of GeV photons at late times ($\gtrsim100\nobreakspace\rm{s}$) or TeV photons at late or early times are challenging synchrotron emission scenarios. An alternative scenario for non-thermal photons at GeV energies is inverse Compton radiation from the external shocks \citep[e.g.][]{bib:HE_GRB_review}.

Determining the highest energies and temporal evolution of GRB spectra thus has important implications for GRB physics and cannot be done by the LAT alone because the effective area is approximately constant above $\approx10\nobreakspace\rm{GeV}$ ($<1\nobreakspace\rm{m^2}$) and the photon flux decreases steeply with energy. Imaging Atmospheric Cherenkov Telescopes (IACTs) are sensitive to gamma rays at very-high energies (VHE; $>100\nobreakspace\rm{GeV}$), however, only upper limits have been reported so far \citep{bib:MAGIC_GRB,bib:HESS_GRB,bib:VERITAS_GRB}. One problem is that IACTs are pointed instruments that need to slew to the GRB position and will therefore in general miss the prompt and early afterglow phase \citep[note, however,][]{bib:HESS_GRB_PROMPT}. The High Altitude Water Cherenkov (HAWC) observatory is a VHE gamma-ray extensive air shower (EAS) detector currently under construction. Its large instantaneous field of view ($\sim2\nobreakspace\rm{sr}$ or 16\% of the sky), near 100\% duty cycle and the lack of observational delays will allow observations during the prompt GRB phase.

VHE photons are attenuated due to interactions with the extra-galactic background light (EBL) and are thus only possible for a very bright and nearby burst. GRB\nobreakspace130427A was an exceptionally bright and nearby GRB which made it a promising target for VHE observations. In this paper, the results of the analysis of HAWC data for this burst are reported.

\section{GRB\nobreakspace130427A}
The prompt phase of GRB\nobreakspace130427A triggered the \emph{Fermi} Gamma-ray Burst Monitor (\emph{Fermi}-GBM) at 07:47:$06.42\nobreakspace\rm{UTC}$ \citep{bib:GRB130427A_Fermi_Science}, denoted $T_0$ in the following. The \emph{Swift} Burst Alert Telescope (BAT) did not trigger immediately, because triggering was de-activated during the slewing to a pre-planned target \citep{bib:GRB130427A_Swift_Science}. GRB\nobreakspace130427A also triggered MAXI/GSC \citep{bib:GRB130427A_MAXI}, SPI-ACS/\emph{INTEGRAL} \citep{bib:GRB130427A_INTEGRAL}, Konus-\emph{Wind} \citep{bib:GRB130427A_Konus_Wind}, \emph{AGILE} \citep{bib:GRB130427A_Agile} and \emph{RHESSI} \citep{bib:GRB130427A_RHESSI}.

The BAT light curve can be divided into three main episodes. First, an initial pulse peaked at $T_0+0.5\nobreakspace\rm{s}$ with a smaller pulse at $T_0+1.1\nobreakspace\rm{s}$. The second, main emission episode starts gradually at $T_0+2.2\nobreakspace\rm{s}$ with a sharp pulse at $T_0+5.4\nobreakspace\rm{s}$ and is followed by a complex structure of various emission peaks lasting a total of about $5\nobreakspace\rm{s}$. There are a few less intense pulses on top of the decay of the main episode, the last pulse peaking at about $T_0+26$~s. A third, much weaker episode starts at $T_0+120\nobreakspace\rm{s}$, with two overlapping pulses peaking at $T_0+131\nobreakspace\rm{s}$ and $T_0+141\nobreakspace\rm{s}$. The decaying emission was detectable until the end of the observations at $T_0+2021$~s. The central time interval, in which 90\% of the prompt flux is detected ($T_{90}$, calculated over the first 1830~s of BAT data) is $(276\pm5)$~s.

At $T_0$, the burst was well within the LAT field of view at a boresight of $47\degr.3$ \citep{bib:GRB130427A_Fermi_Science}. The GBM initiated an Autonomous Repoint Request that started slewing at $T_0+33\nobreakspace\rm{s}$ and brought the burst within $20\degr.1$ of the LAT boresight. The GBM light curve is similar to the BAT one, yielding a $T_{90}$ of $138\nobreakspace\rm{s}$ \citep{bib:GRB130427A_Fermi_GBM}. The LAT Low Energy (LLE, $>10\nobreakspace\rm{MeV}$) emission between $T_0+4\nobreakspace\rm{s}$ and $T_0+12\nobreakspace\rm{s}$ is roughly correlated with the GBM emission. At higher energies (HE, $>100\nobreakspace\rm{MeV}$), there seems to be little correlation with the LLE or GBM emission, beyond an initial spike at $T_0$. The GeV emission is delayed and between $T_0+11.5\nobreakspace\rm{s}$ and $T_0+33\nobreakspace\rm{s}$ an additional power-law component is required in the spectral fit.

\emph{Fermi}-LAT detected temporally extended HE emission from this burst until it became occulted by the Earth after $715\nobreakspace\rm{s}$ \citep{bib:GRB130427A_Fermi_Science}. The burst emerged from occultation at $T_0+3135\nobreakspace\rm{s}$ and remained detectable for about 20~hr (interrupted by further occultations). This unprecedented long HE emission easily surpasses the prominent GRB\nobreakspace940217, where the emission might have lasted more than $5000\nobreakspace\rm{s}$ \citep{EGRET_long_burst}. The GRB fluence is the highest ever detected by the GBM and LAT. Since the GRB redshift was found to be $z=0.3399\pm0.0002$ \citep{bib:redshift_gemini,bib:redshift_NOT,bib:redshift_VLT}, this yields, assuming a standard cosmology, an isotropic energy release of about $8.5\times10^{53}\nobreakspace\rm{erg}$, making it the most energetic GRB so far detected at a redshift $z<0.5$ \citep{bib:Inverse_compton_1}. This burst also featured the most energetic GRB photon ever detected ($95.3\nobreakspace\rm{GeV}$ or $128\nobreakspace\rm{GeV}$ when corrected for the redshift) at $T_0+243\nobreakspace\rm{s}$ \citep{bib:Inverse_compton_2}. This photon appears to be inconsistent with lepton synchrotron radiation in the standard afterglow model \citep{bib:GRB130427A_Fermi_Science}. \cite{bib:synchrotron} find good agreement with the synchrotron model from optical to GeV energies and therefore suggest significant modifications to the relativistic shock physics.

\section{The HAWC Gamma-ray Observatory}
HAWC is an EAS detector currently under construction at Sierra Negra, Mexico, at an altitude of $4100\nobreakspace\rm{m}$ above sea level \citep{bib:HAWC}. It utilises the water Cherenkov technique, where gamma rays are detected by measuring Cherenkov light from secondary particles in an EAS. HAWC has an order of magnitude better sensitivity, angular resolution and background rejection than its predecessor, the Milagro experiment \citep[e.g.][]{bib:Milagro}. Once completed, HAWC will consist of 300 steel tanks of $7.3\nobreakspace\rm{m}$ diameter and $5.0\nobreakspace\rm{m}$ hight, containing a light-tight bladder holding about 188,000 litres of filtered water. Each tank will have three $8^{\prime\prime}$ photomultiplier tubes (PMTs) and one $10^{\prime\prime}$ PMT on the bottom.

HAWC has two data acquisition (DAQ) systems. The main DAQ reads out full air-shower events by recording the time and charge of individual PMT pulses. The signal arrival time in different tanks permits reconstruction of the direction of the incident shower. The scaler DAQ counts the signals in each PMT in $10\nobreakspace\rm{ms}$ windows, which is the finest possible granularity of a light curve. GRBs are detected by a statistical excess over the noise rate \citep[``single particle technique'';][]{bib:scaler_method}. The DAQs complement each other, since they have different energy sensitivities \citep{bib:HAWC_GRBs}.

At $T_0$, data were being collected by the scaler DAQ only using 29 operational tanks out of the 300 planned (called HAWC-30 and HAWC-300 respectively) and 115 deployed PMTs. The data of three PMTs are not considered in the following due to anomalous count rates.

\subsection{Sensitivity}
The zenith angle of GRB\nobreakspace130427A in the HAWC field of view was $57\degr$ and setting at $T_0$. Seven hundred million gamma-ray air showers coming from the direction of GRB\nobreakspace130427A are simulated using CORSIKA version 7.4000 \citep{bib:CORSIKA} with FLUKA version 2011.2b.6 \citep{bib:FLUKA_1,bib:FLUKA_2}. The energy range is $0.5\nobreakspace\rm{GeV}$--$1\nobreakspace\rm{TeV}$, because lower energy showers produce no secondaries that reach the detector and absorption due to EBL cuts off the spectrum at energies beyond that. For comparison, one hundred million gamma-ray air showers randomly distributed on the sky down to a zenith angle of $60\degr$ and energies between $0.5\nobreakspace\rm{GeV}$ and $10\nobreakspace\rm{TeV}$ are simulated.

The detector response is simulated using a detailed detector description within GEANT4 version 10.0.1 \citep{bib:GEANT4_1,bib:GEANT4_2}. It was originally developed for Milagro \citep{bib:Milagro_simulation} and includes detailed descriptions of the geometrical and optical properties of the HAWC tanks and PMTs. The effective area of the scaler system is calculated by counting the hits in all PMTs \citep[for details see][]{bib:HAWC_GRBs}. Since each shower can create hits in multiple PMTs, the effective area is not restricted to the physical size of the detector. The biggest changes in the simulation compared to the previous publication are the inclusion of the fourth, central PMT and a more realistic simulation of the detector (see the discussion of systematic uncertainties below).

\begin{figure}[t]
  \centering
  \plotone{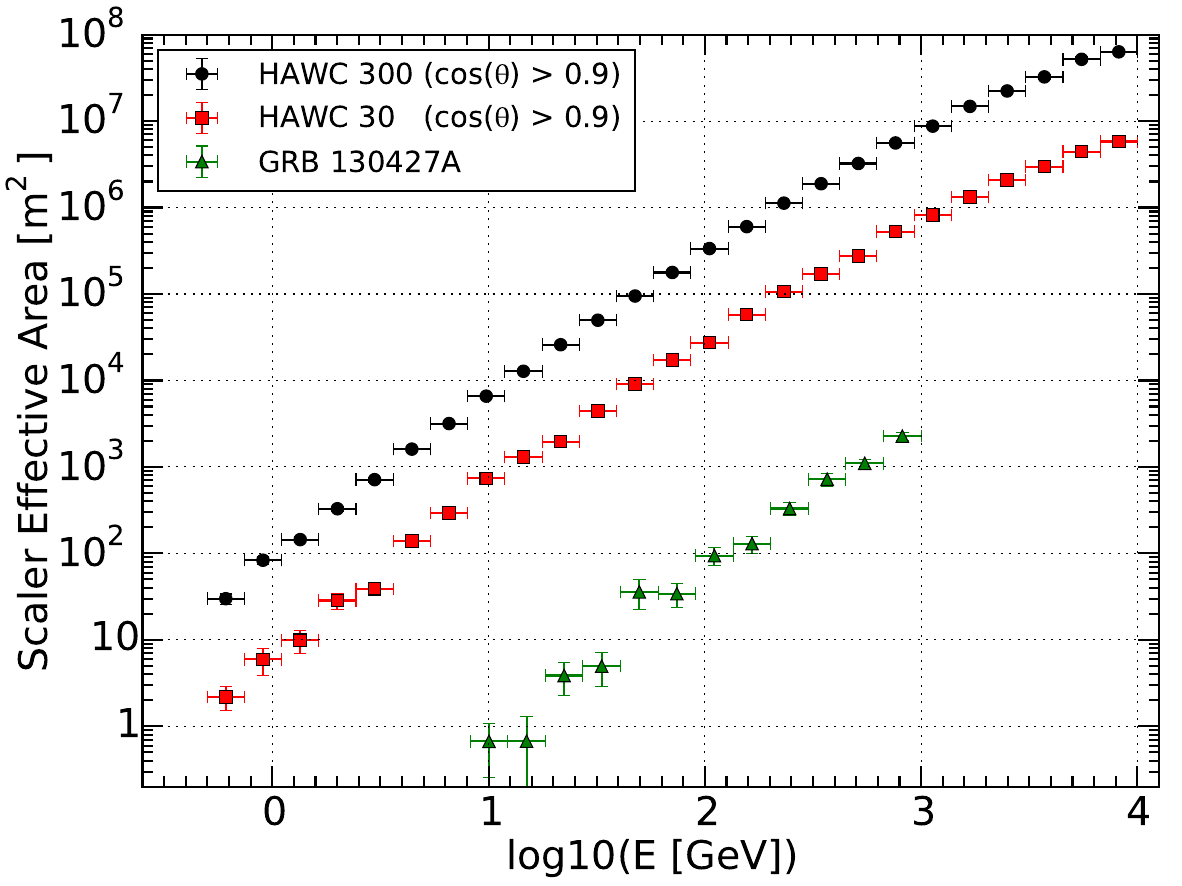}
  \caption{Effective area of the scaler system for HAWC-300 (black dots) and HAWC-30 (red boxes) for showers where the cosine of the zenith angle $\theta$ is larger than 0.9. The green triangles show the effective area for HAWC-30 for showers coming from the direction of GRB\nobreakspace130427A.}
  \label{fig:scaler_effective_area}
\end{figure}

\begin{sloppypar}
Figure\nobreakspace\ref{fig:scaler_effective_area} shows the effective area of the scaler system for HAWC-30 and the direction of GRB\nobreakspace130427A. For comparison, the effective area to bursts overhead for HAWC-30 and HAWC-300 is shown. The comparison between HAWC-30 and HAWC-300 shows that the effective area scales, as expected, linearly with the number of channels. The high zenith angle of GRB\nobreakspace130427A reduces the effective area by more than two orders of magnitude with respect to the zenith. Furthermore, the energy threshold is significantly worsened towards the horizon.
\end{sloppypar}

The width $\sigma_\mu$ of the summed scaler count distribution is larger than that of a Poissonian with mean $\mu$ because correlated sources of noise may produce two or more PMT signals, leading to multiple counting \citep{bib:HAWC_GRBs}. The widening is characterised by the Fano factor $F$: $\sigma_\mu^2=F\mu$. Ten~minutes of data are rebinned into $200\nobreakspace\rm{ms}$ bins that make the count distribution of each channel Gaussian and then $\mu$ and $\sigma_\mu$ are calculated. The mean of the implied Fano factor for a day of data is 12.0. The analysis sensitivity is degraded by $\sqrt{F}$ compared to a purely Poissonian background.

\subsection{Systematic Uncertainties}
\subsubsection{Atmosphere}
The atmosphere has a direct influence on shower development. To model the atmospheric conditions for GRB\nobreakspace130427A, data provided by the Global Data Assimilation System\footnote{http://ready.arl.noaa.gov/gdas1.php} (GDAS) from the closest grid point to the HAWC site are used. Using data from 2013-04-27T09:03:00~UTC, a profile of atmospheric depth is created following the procedure outlined in \cite{bib:atmospheric_profiles}. The deviation to the profile measured three hours earlier is $< 0.5\%$ at each altitude and therefore it can be concluded that the atmosphere was very stable during the observation period and any systematic uncertainty negligible.

For the simulation where the gamma-ray air showers are randomly distributed on the sky, a profile of atmospheric depth is created by averaging the GDAS data for the years 2011 and 2012 at each height. To investigate the systematic uncertainties, two additional profiles that represent the extreme, cumulative deviation from the average profile (above the altitude of HAWC) are selected, which should bracket the systematic error for the air shower development. Calculating the ratio of the effective areas for the two profiles and creating a weighted average over all energies, it was found that the difference is $\sim8\%$. The effective area obtained from the average profile lies in the middle between the two profiles.

\subsubsection{Water Quality}
The propagation of Cherenkov light is influenced by the water in the tanks. The attenuation length has been measured at $405\nobreakspace\rm{nm}$ using $1\nobreakspace\rm{m}$ long water samples taken from 26 tanks and varies between $5\nobreakspace\rm{m}$ and $16\nobreakspace\rm{m}$. Since all tanks contribute equally to the scaler effective area (e.g. two tanks with 50\% more and less than the average counts give the same summed counts as two average tanks), the attenuation length can be averaged over all tanks. Taking into account the number of PMTs per tank considered in the analysis, this yields an attenuation length of $10.4\nobreakspace\rm{m}$. Attenuation can be caused by absorption or scattering and these processes cannot be disentangled with current measurements. Absorption has a bigger impact on the scaler effective area, because a scattered photon may still be detected. The bladder holding the water is black on the inside to avoid reflection of photons. GEANT4 simulations are used to determine absorption and scattering parameters compatible with the measured attenuation length.

Vertically down-going muons provide well defined charge and timing signals in a HAWC tank and can be selected from raw data by applying appropriate charge and timing cuts. First studies indicate that the Monte Carlo describes the data well, assuming scattering properties comparable to those of clear seawater \citep{bib:water_scattering} and an absorption spectrum for water \citep{bib:water_absorption} scaled so that the combination of absorption and scattering is compatible with the measured attenuation length at $405\nobreakspace\rm{nm}$. The same water properties have been used in the simulation where the gamma-ray air showers are randomly distributed on the sky.

The absorption spectrum between $300\nobreakspace\rm{nm}$ and $550\nobreakspace\rm{nm}$ varies strongly depending on impurities in the water. To characterise the systematic uncertainty, two simulations were done, with scattering either highly effective and causing most of the measured attenuation or highly ineffective and requiring more absorption. Calculating the ratio of the effective areas and creating a weighted average over all energies, it was found that the difference is 19\%, which quantifies the total estimate for the systematic uncertainties, and the effective area given here for GRB\nobreakspace130427A is the average of these two extreme cases.

\section{Data Analysis}
\subsection{Search Window Selection}
A transient flux of gamma rays is identified by searching for an excess of the summed PMT counts over an expected background in a certain search window (SW). A first set of six partially overlapping SWs has been selected based on GRB Coordinates Network\footnote{http://gcn.gsfc.nasa.gov} (GCN) notices. The first window from $T_0+0\nobreakspace\rm{s}$ to $T_0+20\nobreakspace\rm{s}$ covers the first and second main emission periods seen by BAT and the second window ($T_0-5\nobreakspace\rm{s}$ to $T_0+55\nobreakspace\rm{s}$) is an extension of that. The $T_{90}$ reported by GBM motivated the third time window ($T_0-5\nobreakspace\rm{s}$ to $T_0+145\nobreakspace\rm{s}$). A time window from $T_0+120\nobreakspace\rm{s}$ to $T_0+300\nobreakspace\rm{s}$ was selected to cover the third emission period seen by BAT. All three BAT emission periods were combined in a time window from $T_0-10\nobreakspace\rm{s}$ to $T_0+290\nobreakspace\rm{s}$. Additionally, a time window $-10\nobreakspace\rm{s}$ to $10\nobreakspace\rm{s}$ around the time of the highest energy LAT photon was chosen. Preliminary results were reported in \cite{bib:HAWC_GCN}. Besides the SWs, the data between $-1200\nobreakspace\rm{s}$ and $+1200\nobreakspace\rm{s}$ around the GRB trigger time were kept blind for future analysis.

A second set of SWs not previously reported is selected using intervals with hard power-law components in the GeV light curve. These are the window from $T_0+11.5\nobreakspace\rm{s}$ to $T_0+33\nobreakspace\rm{s}$ \citep[index $-1.66$,][]{bib:GRB130427A_Fermi_Science} and the time window from $T_0+196\nobreakspace\rm{s}$ to $T_0+257\nobreakspace\rm{s}$, where the power-law fit has a spectral index harder than $-2$.

\subsection{Background Estimation}\label{subsec:background_estimate}
\begin{figure}[t]
 \centering
 \plotone{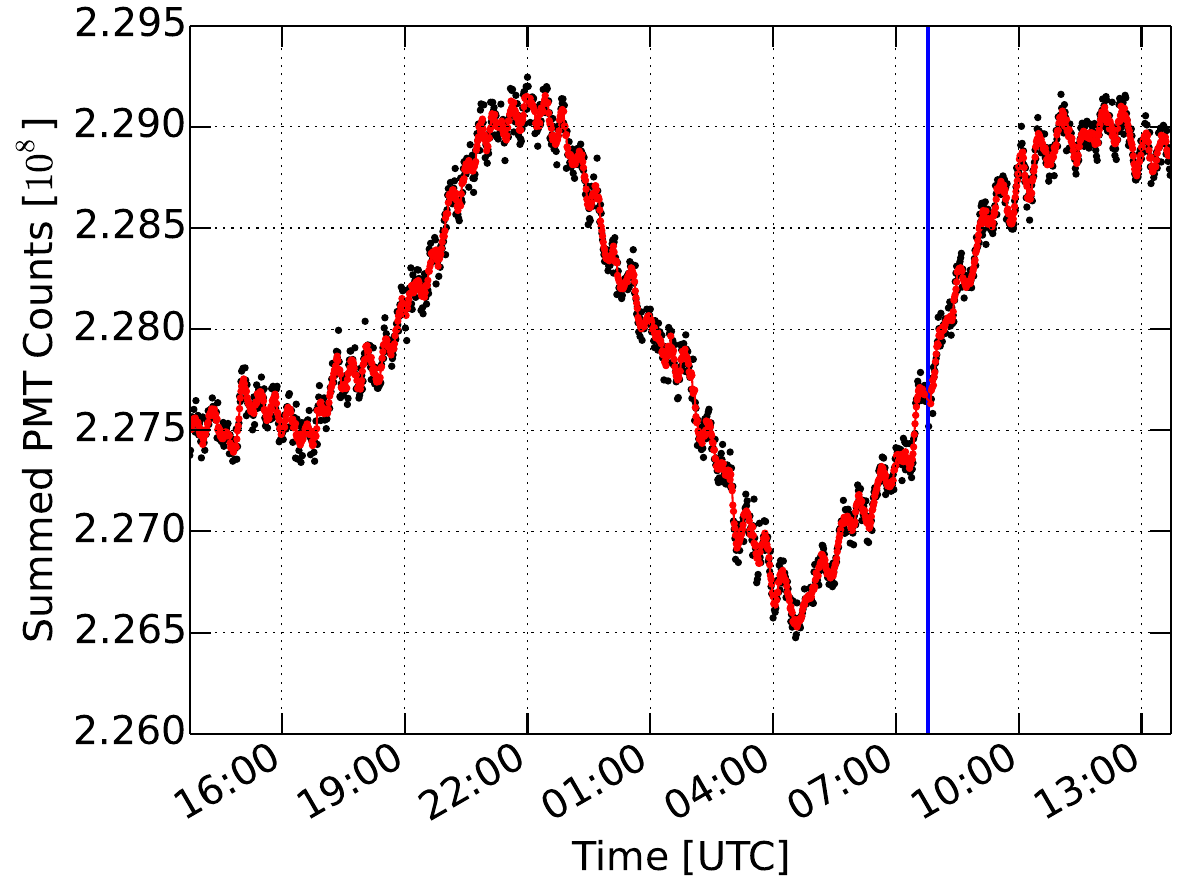}
 \caption{Black dots show the summed counts of all selected PMT channels (GRB\nobreakspace130427A trigger time shown as blue line) using a $60\nobreakspace\rm{s}$ binning. The data are well described by the moving average (red line).}
 \label{fig:GRB130427A_data}
\end{figure}

The scaler data is rebinned to the different SW sizes. Figure\nobreakspace\ref{fig:GRB130427A_data} shows an example of the summed count rates of all selected PMT channels rebinned to $60\nobreakspace\rm{s}$ (black points). The rates are influenced at the percent level by daily changes in the atmospheric pressure and temperature leading to differences in the shower development. A clear 12~hr modulation can be seen that corresponds to the local pressure cycle. This is long compared to the SWs and makes it unnecessary to correct the data for these environmental effects. The rates are also influenced by the temperature of the electronics, which has a $\sim$27~minute cycle that can be seen in the data as well.

For each rebinning, a symmetric moving average (MA) is applied to each channel, where each point $i$ is replaced by the average of the $N$ points before and after (each having number of counts $C$):
\begin{equation}\label{eq:moving average}
  {\rm MA}(i) = \frac{1}{2N} \sum^{j=i+N}_{j=i-N;j \neq i} C(j)
\end{equation}

\begin{figure}[t]
 \centering
 \plotone{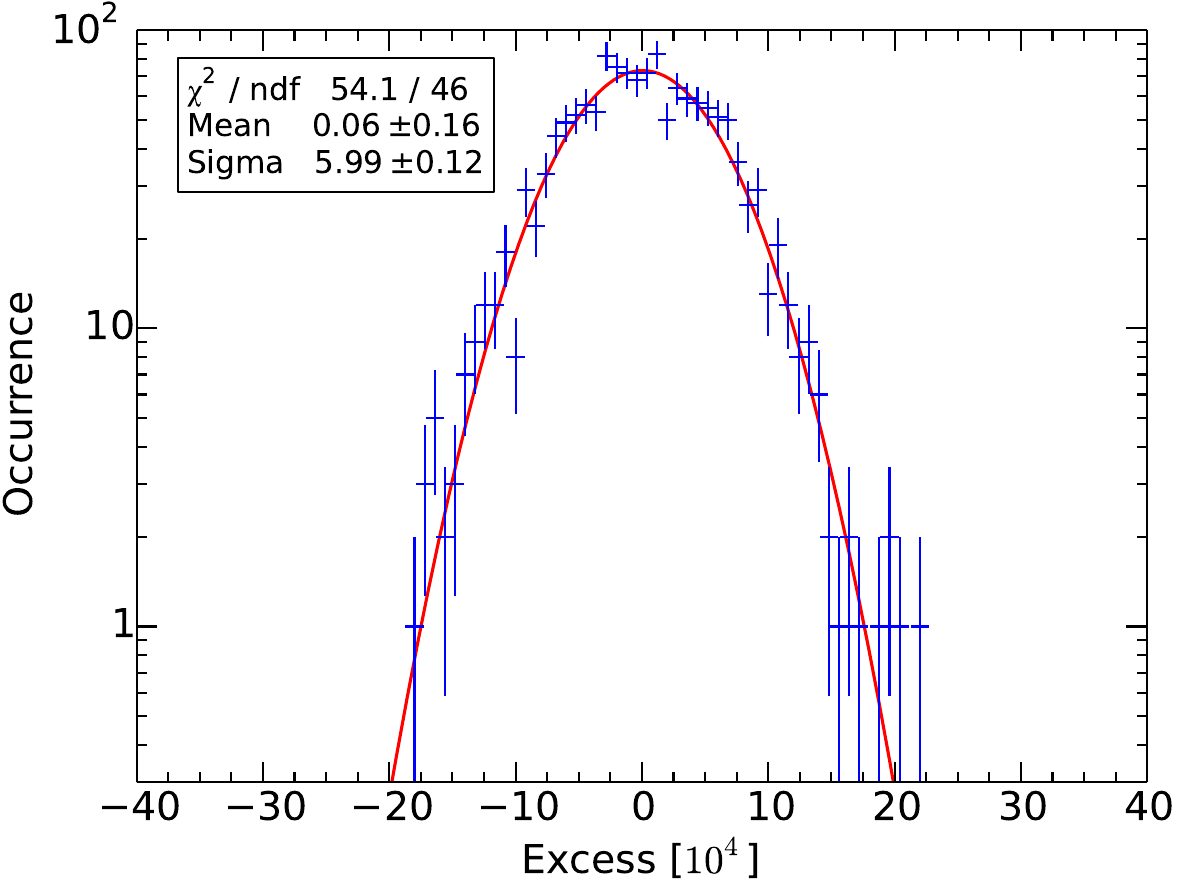}
 \caption{Histogram of the residuals between data and the moving average in the $60\nobreakspace\rm{s}$ binning. A Gaussian fit (red line) describes the data well.}
 \label{fig:GRB130427A_residuals}
\end{figure}

The red line in Figure\nobreakspace\ref{fig:GRB130427A_data} shows the summed MAs. The differences to the data yield an excess distribution (see Figure\nobreakspace\ref{fig:GRB130427A_residuals}), which shows no significant outliers and is well fit by a Gaussian (also true for the other binnings used). The analysis is more sensitive, the narrower this distribution is. An optimal sensitivity is obtained when $N$ corresponds to an interval of 3~minutes on each side. For the $20\nobreakspace\rm{s}$ and $21.5\nobreakspace\rm{s}$ binning, $N=9$ is used and for $60\nobreakspace\rm{s}$ and $61\nobreakspace\rm{s}$ $N=3$. Comparing the mean of the MA points to the width of the Gaussian fitted to the excess distribution implies Fano factors (see Table\nobreakspace\ref{tab:results}) close to the previously derived value. For all larger binnings, $N=1$ is chosen. The Fano factor increases drastically in this case, reducing the sensitivity for the longest time window by more than a factor of two. Apparently, the MA is not optimal for longer time windows, possibly because of the cooling cycle on a similar timescale. This is a consideration for future enhancement of the analysis method.

Some points of the MA contain data from the SWs and would thus bias the background estimation. Those points are excluded and an unbiased background estimate for each SW is obtained by averaging the last and first point outside those times. Additionally, these points are also excluded from the excess distributions. The $p$-value of the excess in the SW, which is the probability that the observed or a higher excess is caused only by background, is calculated using the excess distribution.

\section{Results}
\begin{deluxetable}{l|cccccccc}
\tabletypesize{\footnotesize}
\tablecolumns{8}
\tablewidth{0pt}
\tablecaption{Results of the HAWC Analysis\label{tab:results}}
\tablehead{\vspace{-0.1cm} & \colhead{PMT Sum}  & \colhead{BG ESt.}  & \colhead{Excess}   & \colhead{$p$-value} & \colhead{Upper Limit} & \colhead{Sensitivity} & \\
           \vspace{-0.1cm} & \colhead{}  & \colhead{}  & \colhead{}   & \colhead{} & \colhead{} & \colhead{} & \colhead{Fano Factor} \\
           & \colhead{$(10^4\nobreakspace\rm{counts})$} & \colhead{$(10^4\nobreakspace\rm{counts})$} & \colhead{$(10^4\nobreakspace\rm{counts})$} & \colhead{(\%)}    & \colhead{$(10^4\nobreakspace\rm{counts})$}    & \colhead{$(10^4\nobreakspace\rm{counts})$}}
\startdata
  0--20~s    &   7593.0           &   7589.0           &   4.0              & 10.1              &  9.27                 &  5.46 & 13\\
233--253~s   &   7590.6           &   7589.7           &   0.9              & 39.0              &  6.06                 &  5.46 & 13\\
11.5--33~s   &   8161.3           &   8157.8           &   3.4              & 15.0              &  8.85                 &  5.69 & 13\\
 $-5$--55~s  &  22765.5           &  22765.4           &   0.2              & 48.9              & 10.15                 & 10.46 & 16\\
196--257~s   &  23148.1           &  23148.2           &  $-0.1$            & 50.1              &  9.87                 & 10.50 & 16\\
 $-5$--145~s &  56899.3           &  56927.5           & $-28.2$            & 98.9              &  3.84                 & 20.32 & 25\\
120--300~s   &  68308.5           &  68325.9           & $-17.4$            & 90.3              &  9.15                 & 23.57 & 28\\
$-10$--290~s & 113826.7           & 113895.5           & $-68.8$            & 99.6              & 10.96                 & 52.08 & 77\\
\enddata
\tablecomments{For each search window, the sum of all PMT counts in that window is given together with the background estimation from the MA. From these two values the excess is calculated (using unrounded numbers). The $p$-value gives the probability that the background produces an equal or higher excess than the one observed (not accounting for the different trials). The event upper limits correspond to a 90\% confidence level. Sensitivity is the corresponding average upper limit as defined in \cite{bib:Feldman_Cousins}. The Fano factor is calculated as described in Section\nobreakspace\ref{subsec:background_estimate}.}
\end{deluxetable}

Table~\ref{tab:results} shows the results for each SW. All are consistent with the assumption of background only. Using Gaussian fits of the excess distributions, 90\% confidence belts are constructed using the method described by \cite{bib:Feldman_Cousins} and upper limits on the number of excess events are derived. These upper limits are converted to integral flux upper limits between $0.5\nobreakspace\rm{GeV}$--$1\nobreakspace\rm{TeV}$ using the HAWC effective area for GRB\nobreakspace130427A.

\begin{figure}[t]
 \centering
 \plotone{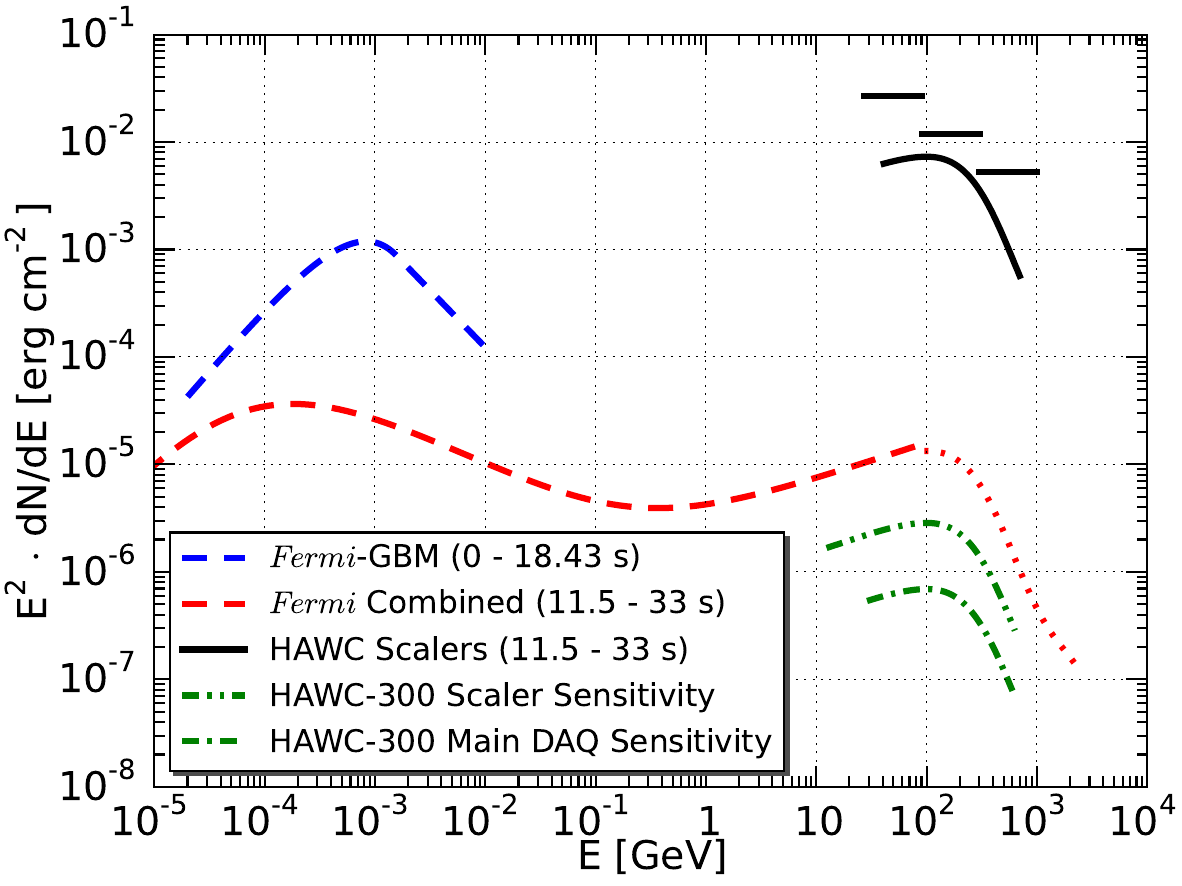}
 \caption{Dashed lines show the spectra fitted to the prompt GBM data (small energy range, blue) and the combined fit of GBM and LAT data when an additional power law is significant (larger energy range, red). The dotted continuation shows the effect of EBL absorption. Black lines show the scaler limit as an integral upper limit assuming the \emph{Fermi} fit plus EBL absorption and as ``quasi-differential'' limits assuming $E^{-2}$. The green dashed--dotted lines show the sensitivity of the two HAWC DAQs for an overhead GRB. The limits are restricted to the energy range of 90\% of the expected counts.}
 \label{fig:limits}
\end{figure}

Figure\nobreakspace\ref{fig:limits} visualises the $11.5$--$33\nobreakspace\rm{s}$ limit. The additional power law reported by LAT becomes significant after the GBM detected emission has faded and appears temporally distinct, suggesting that it might arise from a different emission region or mechanism \citep{bib:GRB130427A_Fermi_Science}. Assuming no intrinsic cut-off and modelling the absorption on the EBL according to the fiducial model in \cite{bib:EBL}, the combined best fit of GBM and LAT data is extrapolated to VHEs (dotted line). Under this assumption, the cut-off due to EBL happens just beyond the end of the LAT energy coverage. The scaler limit is calculated assuming the dotted line as the spectral shape. Since the scaler DAQ has no energy information, ``quasi-differential'' limits are calculated by restricting an $E^{-2}$ spectrum to the energy ranges indicated by the limit bars. The green dashed--dotted lines show the sensitivity of the two HAWC DAQs for the full detector for an overhead GRB. For the main DAQ, the line indicates the flux level which leads to a 50\% probability for detecting a $5\sigma$ excess. For the scaler DAQ, the line shows the expected average upper limit. Since the Fano factor increases with the number of PMTs (and thereby reduces the sensitivity of the analysis), the scaler limit has been scaled by a factor of $\sqrt{3}$, assuming that the Fano factor increases by a factor of three in HAWC-300.

\section{Discussion}
HAWC provides the first limits on the prompt emission of GRB\nobreakspace130427A in the VHE range. Due to the high zenith angle of the GRB and the incomplete HAWC detector, the limits are about two orders of magnitude higher than a simple extrapolation of the \emph{Fermi} data. Had the GRB been observed overhead (cosine of zenith angle larger than 0.9), then it would have produced a $1.9\sigma$ effect in the HAWC-30 scaler DAQ. A similar GRB close to zenith would be easily detected by the full HAWC detector, which will be completed soon. In addition to exceptional bursts like GRB\nobreakspace130427A, HAWC can also detect other GRBs with a rate as high as 1--2 GRBs per year \citep{bib:HAWC_GRB_rate}.

\cite{bib:Inverse_compton_2} showed that the LAT emission of GRB\nobreakspace130427 in five different time intervals fits well to a power law with an index of $\Gamma\sim-2$. Evidence was presented that after $3000\nobreakspace\rm{s}$ a broken power-law model with break energy $\sim1\nobreakspace\rm{GeV}$, a soft low-energy component ($\Gamma\sim-2.6$) and a hard high-energy component ($\Gamma\sim-1.4$) is preferred at the $2.9\sigma$ level. The latter has properties consistent with synchrotron self-Compton (SSC) emission \citep{bib:Inverse_compton_1,bib:Inverse_compton_2, bib:Inverse_compton_3}. A second analysis found that the GRB spectrum is well described by a power law at all times and a broken power law is not statistically required \citep{bib:GRB130427A_Fermi_Science}. A second broken power-law described by \cite{bib:Inverse_compton_2} between $138\nobreakspace\rm{s}$ and $750\nobreakspace\rm{s}$ was explained by \cite{bib:GRB130427A_Fermi_Science} as an effect of power laws with varying spectral indices over time. The VERITAS array has observed GRB\nobreakspace130427 $\sim20\nobreakspace\rm{hr}$ after the onset of the burst \citep{bib:VERITAS}. The data disfavour the interpretation that the LAT emission originates from the SSC mechanism. Observations of GRB\nobreakspace130427A with HAWC-300 at a favourable zenith angle would have provided clear evidence for or against a possible SSC emission scenario.

HAWC can observe the prompt phase of GRBs that are typically not accessible to IACTs due to observational delays. A possible light curve could probe synchrotron and SSC scenarios and might reveal information on the interstellar density and magnetic field of the surrounding medium \citep[e.g.][]{bib:short_bursts_VHE}. A spectrum at TeV energies could also probe possible hadronic-dominated scenarios, e.g. due to proton synchrotron emission \citep{bib:proton_synchrotron}. A joint main and scaler DAQ analysis could provide information on a possible cut-off beyond the LAT energy range \citep{bib:HAWC_GRBs}. Such a cut-off could be the signature of gamma ray absorption, either in the GRB (thus providing a probe of the bulk Lorentz factor) or along the line of sight (thus providing a probe of the EBL), or may also indicate the maximum particle energy produced by the GRB.

\begin{acknowledgements}
We acknowledge the support from:
the US National Science Foundation (NSF);
the US Department of Energy Office of High-Energy Physics;
the Laboratory Directed Research and Development (LDRD)
program of Los Alamos National Laboratory;
Consejo Nacional de Ciencia y Tecnolog\'{\i}a (CONACyT), Mexico (grants 55155, 105666, 122331, 132197);
Red de F\'{\i}sica de Altas Energ\'{\i}as, Mexico;
DGAPA-UNAM (grants IG100414-3, IN108713, IN121309, IN115409, IN113612);
VIEP-BUAP (grant 161-EXC-2011);
the University of Wisconsin Alumni Research Foundation;
the Institute of Geophysics, Planetary Physics, and
Signatures at Los Alamos National Laboratory;
the Luc Binette Foundation UNAM Postdoctoral Fellowship program.
\end{acknowledgements}

\bibliographystyle{apj}
\bibliography{references}


\end{document}